# Hysteretic giant magnetoimpedance effect analyzed by first-order reversal curves

**F. Béron[1,\*], L. A. Valenzuela[1], M. Knobel[1], L. G. C. Melo[2] and K.R. Pirota[1]**

[1]Instituto de Física Gleb Wataghin, Universidade Estadual de Campinas, 13083-970, Campinas (SP), Brazil

[2]Applied Electromagnetics Group, Universidade Federal de São João del-Rei, 36420-000, Ouro Branco (MG), Brazil

## Abstract

Hysteretic giant magnetoimpedance (GMI) of amorphous ribbons with a well-defined transversal domain structure is investigated by means of first-order reversal curves (FORC) analysis. The FORCs are not confined to the hysteretic area, exceeding the major curve amplitude. Irreversible switches of the transverse permeability, caused by domain wall structure transitions, may be the origin of the observed FORC distribution. An interlinked hysteron/anti-hysteron model is proposed to interpret it, which allows analyzing the influence of frequency and magnetostriction upon the hysteretic GMI effect.

**I. INTRODUCTION**

The giant magnetoimpedance (GMI) effect consists of a drastic change (up to hundreds of percent) of the electrical impedance $Z$ of a magnetically soft conductor upon application of an external magnetic field $H$. It is related to variations of the effective magnetic permeability $\mu$, which is strongly affected by $H$ and the frequency of the ac driving current, $f$, and that governs the so-called penetration depth through the sample. Although it was first observed around seven decades ago [1], its intense investigation started only in 1994 [2]. This "rediscovery" of GMI in amorphous soft magnetic alloys attracted much attention of scientific community owing to its potential application in ultra-sensitive magnetic sensors and reading heads. Since both static and dynamic magnetic behaviors of the material depend on $\mu$, GMI could also be used as a tool of investigation for materials parameters such as saturation magnetization, anisotropy field, magnetostriction coefficient, conductivity, and damping factor, among others. In spite of these possibilities, the exploration of the GMI effect as a characterizing tool has been still rather limited to few exploratory studies [3]. In order to understand the experimental results and to guide the design of materials with enhanced GMI response, several theoretical models have been developed to find an approximate expression for the transversal $\mu$ [4-6]. The commonly employed formalism is based on the simultaneous solution of Maxwell and Landau-Lifshitz equations, whose solutions, or modes, are used to satisfy the boundary conditions for a particular geometry, from which $\mu$ and $Z$ can be obtained [5, 6]. These models explain the GMI behavior in a broad range of $H$ and $f$, although improvements have been necessary to include unsaturated behavior ($H$ less than the anisotropy field $H_k$, about 1 Oe for a typical soft magnetic metal) [7]. One notable example is the hysteretic behavior of the GMI curve, observed for relatively low frequencies (tens of MHz and below) [7-8]. Surprisingly, there are few studies about the hysteretic GMI effect, even considering its technological relevance, especially for applications of GMI at low fields [8]. The hysteresis can be detrimental for some applications, like magnetic sensors, while others can take advantage of it, such as memories. Therefore, from both fundamental and technological viewpoints, the investigation of the hysteretic GMI effect is of primary importance.

The first-order reversal curve (FORC) method is one of the most powerful tools to investigate the origins and to characterize the materials general hysteretic behavior [9]. Since 1985 it has been successfully applied to probe the magnetization ($M$) hysteresis in several systems [10-13], but its mathematical nature allows a more general use [9]. The FORC method has already been successfully used to investigate the hysteretic behavior of other parameters: ferro-electricity [14], pressure [15] and giant magnetoresistance (GMR) [16]. The main advantage over major hysteresis curves is that it gives the distribution of local properties, instead of average, which can be crucial when dealing with nanostructured systems, for example. Thereby, we think that the FORC method should be seen as a powerful experimental tool that can be used to probe the hysteretic behavior of any system that respects the wipe-out and congruency conditions [9].

In this article, the measurement formalism of the FORC method was applied to the GMI response as a function of the external field of FeCoSiB amorphous ribbons with well-defined transversal domain structure. While their GMI responses exhibit hysteresis at low field, the magnetization present a fully reversible behavior. We conceived a new type of hysteron to adapt the traditional FORC analysis to the particular type of hysteresis exhibited by the GMI signal. Following ref. [17], its physical signification is explained in terms of the transitions between Bloch and Néel domain walls. The novel established FORC procedure for GMI signal (called GMI-FORC) allows the investigation of the effects of different parameters (current amplitude and frequency, ribbon characteristics, etc.) on the domain wall transitions fields, and so their specific repercussions on the ribbon magnetic structure.

## II. FORC METHOD

The FORC method is based on the classical Preisach model. The global behavior of the hysteresis is associated with a collection of single square irreversible curves, called mathematical hysterons and representing the hysteresis operators [9]. The method in itself consists of the measurement of increasing minor hysteresis curves starting from different input values, called reversal points ($H_r$ for field-driven hysteresis), and reaching the positive saturation. The mathematical hysterons distribution, the so-called FORC distribution function, $\rho(H, H_r)$, is obtained through the calculation of a second-order mixed derivative of the output variable in function of the reversal and actual input values [9].

Compared to the major hysteresis curve, which gives the average behavior of the hysteretic operators, it presents the advantage to yield the function distribution of those. For complex systems, hysteretic phenomena can not be directly modeled as mathematical hysterons, like ribbon's GMI as function of external field, but also most of magnetization reversal mechanisms (domain wall nucleation-propagation, vortex, coherent rotation away from the easy axis, etc.). In these cases, an essential step consists of modeling an assembly of mathematical hysterons that adequately describes the behavior of each hysteretic process occurring. For GMI-FORC, we elaborated a dual-hysteron model that represents the elementary irreversible behavior of the GMI (see Sec. IV). For further interpretation of the FORC diagram, one needs to correctly identify the physical meaning of those dual-hysterons.

## III. RIBBONS CHARACTERIZATION

Amorphous ribbons of $(Fe_xCo_{1-x})_{70}Si_{12}B_{18}$ ($x = 0.045$ - $0.050$), 20 μm thick, were prepared by melt spinning technique. A proper annealing treatment, described in [18], induced magnetostriction, which constant depends on the Fe/Co ratio. It results of a uniaxial anisotropy with easy axis perpendicular to the ribbon axis, as confirmed by a representative magnetic optical Kerr effect (MOKE) image (Fig. 1). The magnetization major curve in function of a longitudinal applied field presents a fully reversible behavior, without detectable hysteresis, as expected due to the transversal anisotropy (Fig. 2).

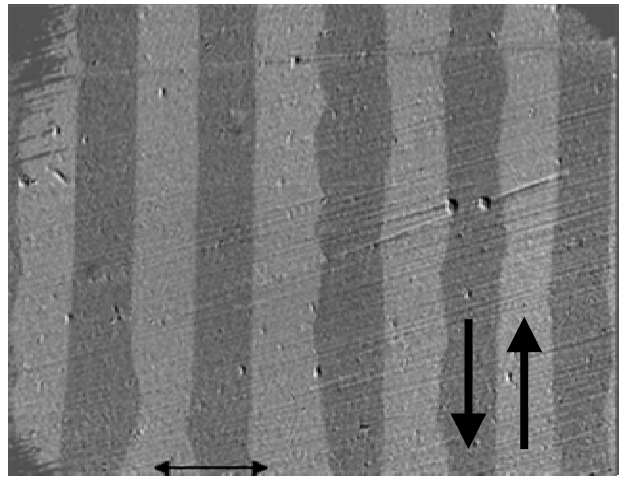

**50 μm**

FIG 1. Top-view MOKE image of the studied FeCoSiB ribbon. The alternation of dark and clear domains signifies antiparallel perpendicular domains, resulting of the well-defined transversal magnetic anisotropy. The arrows indicate the magnetization direction.

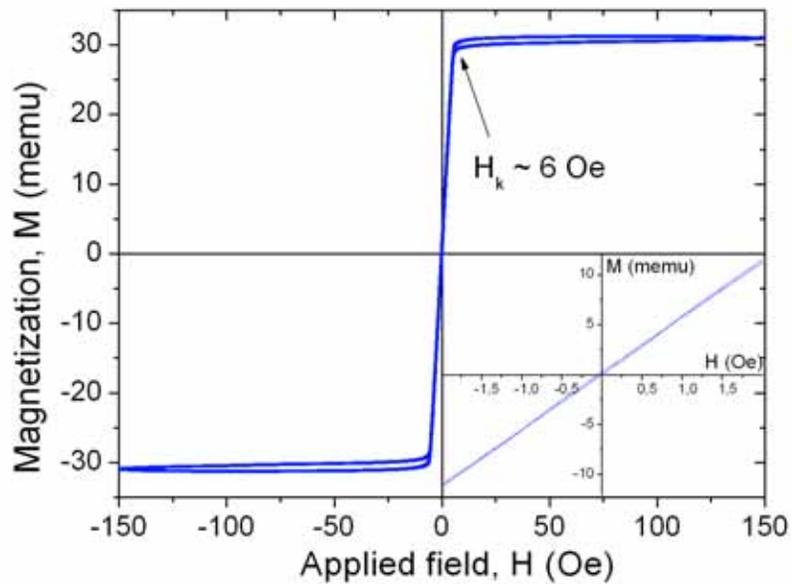

FIG 2. Typical magnetization major curve of FeCoSiB ribbon (longitudinal applied field, $x = 0.045$). Inset: Zoom of the low field region.

For a question of comparison with further results about GMI-FORC, the FORC method was applied on the magnetization curve. Because of the low field involved and the ribbon geometry, a special high-precision AC induction magnetometer was built and adapted to FORC measurement [19]. The obtained FORC diagram does not exhibit any FORC distribution pattern other than noise (Fig. 3), which was expected from the reversible magnetization curve. The scale of the FORC distribution (z axis) was considerably lowered until the noise level in order to clearly show that it only presents a null FORC distribution, i.e. no irreversible processes.

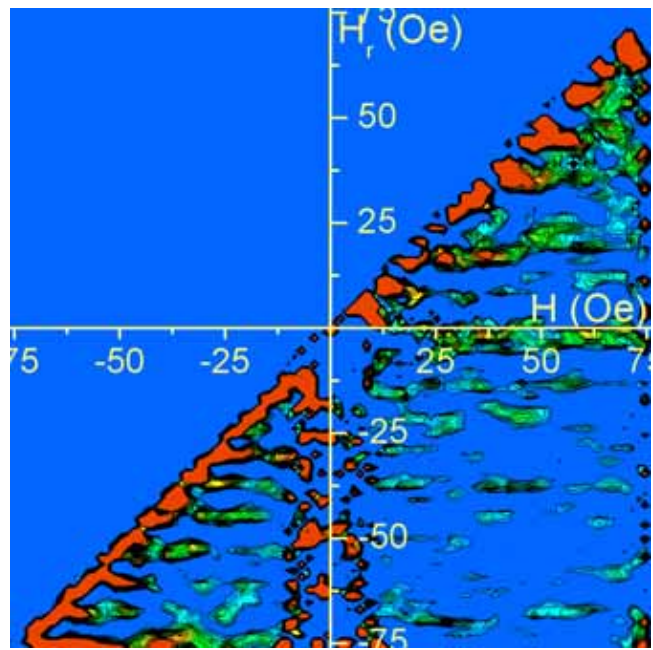

FIG. 3. FORC diagram from the magnetization of FeCoSiB ribbon (longitudinal applied field, saturation field = 160 Oe, reversal field step = 1 Oe, applied field step = 0.5 Oe, $x = 0.045$).

On the other side, the GMI curve as a function of the longitudinal static applied field reveals a different behavior. Figure 4 presents a typical complete GMI signal (real part of $Z$, $R$, ac current amplitude $i = 1$ mA, frequency $f = 500$ kHz, $x = 0.045$) where the two typical peaks associated with the anisotropy field are clearly visible around $H = \pm 6$

Oe. Moreover, it exhibits a hysteretic region at low field ($H = \pm$ 4.5 Oe), which is symmetric about the origin. This discrepancy between the hysteretic behaviors of GMI and magnetization clearly indicates that they could arise from different physical processes, thus exhibiting different behavior as function of applied field. Therefore, GMI-FORCs are probing the irreversibility of the transverse permeability $\mu_\tau$, while magnetization FORCs are probing the irreversibility of magnetization reversal processes. The combination of both could lead to a better understanding of the local processes occurring into soft amorphous ribbons.

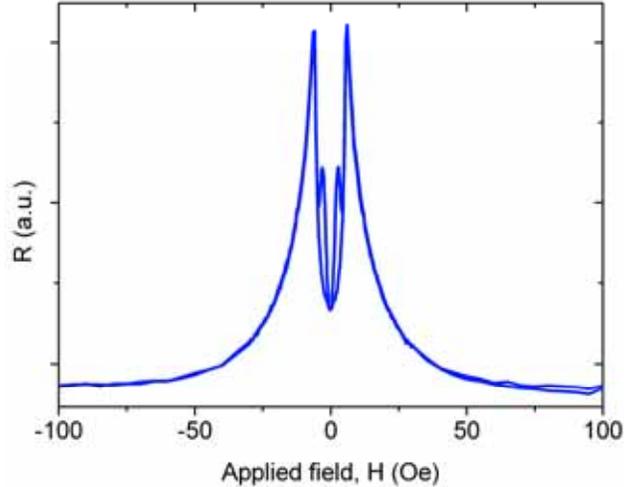

FIG. 4. Typical major curve of the FeCoSiB ribbon GMI signal ($R$, $x = 0.045$, $f = 500$ kHz, $i = 1$ mA).

Domain walls (DWs) creation, annihilation or structure modification represents possible source of irreversible change in the transverse permeability of metallic amorphous ribbons. For thick NiFe films, Middelhoek proposed a simple model based on a magnetic structure formed by either Néel or Bloch DWs, where the DW type depends upon the applied field strength [20]. It was subsequently employed to qualitatively explain the ferromagnetic resonance of unsaturated NiFe thin films [17]. Here, we used the Middelhoek's Bloch-Néel DW model to interpret the hysteretic GMI behavior of soft magnetic ribbons. For the typical ribbon thickness (20-30 μm), in order to minimize the magnetostatic energy, the creation of Bloch walls is expected when the magnetization splits into perpendicular domains. Later, the Néel walls become favorable beyond a certain magnetization critical angle respective to the external field direction. With a domain wall structure transformation, one should not expect a hysteresic behavior of the magnetization, because both Bloch and Néel domain walls present the same magnetization along the longitudinal direction.

From the anhysteretic and linear magnetization major curve (Figure 2) and using typical values of DW energy density given in Ref. 20, one is able to evaluate the magnetic field needed to reach the critical angle for the transition. The expression for the Bloch type domain wall energy density is given by [20]

$$E_B(\theta) = E_B(180)\cos^2\theta \qquad (1)$$

where $\theta$ is the angle that the domains make with the easy magnetization axis when a static field is applied along the hard direction. The component of the magnetization along the easy magnetization direction is therefore $M_s\cos\theta$. For the Néel wall, the energy density is proportional to the square of the component of the magnetization which causes volume charges, $M_s(1-\sin\theta)$, and one can assume:

$$E_N(\theta) = E_N(180)\left[1 - \sin\theta\right]^2 \qquad (2)$$

The transition angle, $\theta_c$, occurs when

$$E_B(180)\cos^2\theta_c = E_N(180)\left[1 - \sin\theta_c\right]^2 \qquad (3)$$

Using Middelhoek's typical values for NiFe-Py films ($E_B(180) = 4.0$ ergs/cm$^2$ and $E_N(180) = 8.5$ ergs/cm$^2$ [20]), we obtain $\sin\theta_c = 0.36$, leading to a value of the longitudinal magnetization (hard direction) of around $M_s\sin\theta_c \sim 0.01$ emu. According to the ribbon magnetization curve (see Fig. 2), this magnetization value corresponds to a longitudinal field of around 2 Oe.

## VI. GMI-FORC

## A.  FORCs

The measured FORC curves (GMI-FORCs), were obtained by measuring the impedance, at a fixed frequency, when $H$ is varied from $H_r$ back to positive saturation (Fig 5, thin colored lines). They were confined in this region, because the $R$ reversible behavior occurring at higher field yields a null FORC distribution.

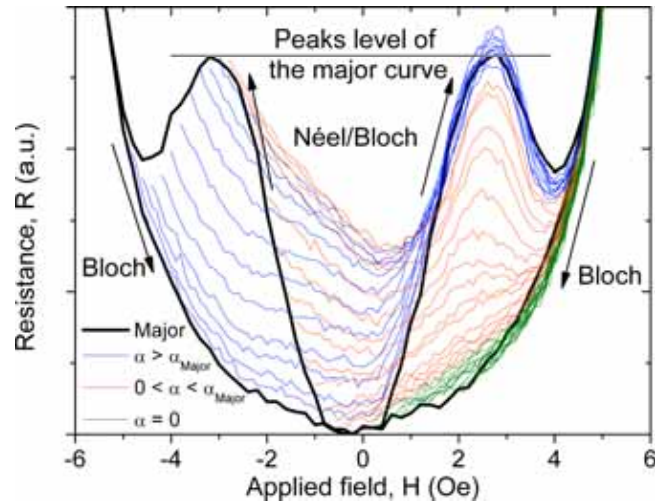

FIG. 5. Low field region of typical major curve (black thick line) and FORCs (thin colored lines, separated into three groups according to their behavior) of the amorphous ribbon GMI signal ($R$, $x = 0.045$, $f = 500$ kHz, $i = 1$ mA). The reversal field step ($\Delta H_r$) and field step ($\Delta H$) were respectively taken as 0.2 and 0.1 Oe, while 100 Oe was applied to completely saturate the ribbon between each FORC.

The first striking observation while looking to the FORCs is that they are not confined within the two paths described by the major curve, passing through the two hysteretic areas as well as outside them (Fig. 5). Therefore, the $\mu_t$ irreversibility of the entire ribbon is not constituted of only two states, but of several distinct states. According to the hypothesis that this irreversibility is provoked by change in the DW structure, the measurement of the GMI-FORCs clearly indicates that the DWs do not undergo a simultaneous structure transition. Intermediate states, i.e. with both Bloch and Néel walls, are energetically possible. This change is consequently thought to be progressive, which can be probed due to the particular FORC measurement. Assuming that no additional DW transition had occurred since $H_r$, corroborated by the similitude of the FORCs path and the major lower branch, where none are expected, remanent GMI signal variation is directly and only proportional to the Néel/Bloch DW ratio α, each structure having a distinct $\mu_t$ ($\mu_{\text{Néel}} > \mu_{\text{Bloch}}$). The behavior of $R_H = 0$ Oe suggests that the newly formed Néel walls progressively switch back into Bloch ones, being minimum where all the DWs are of Bloch type and maximum for $H_r = -2.7$ Oe, indicating the applied field for which the system reaches the highest α (Fig. 6). It's worth noting that it does not correspond with the GMI peak position (±3.2 Oe). Further studies are necessary to quantify α, which could be achieved by calculating the respective $\mu_t$ values.

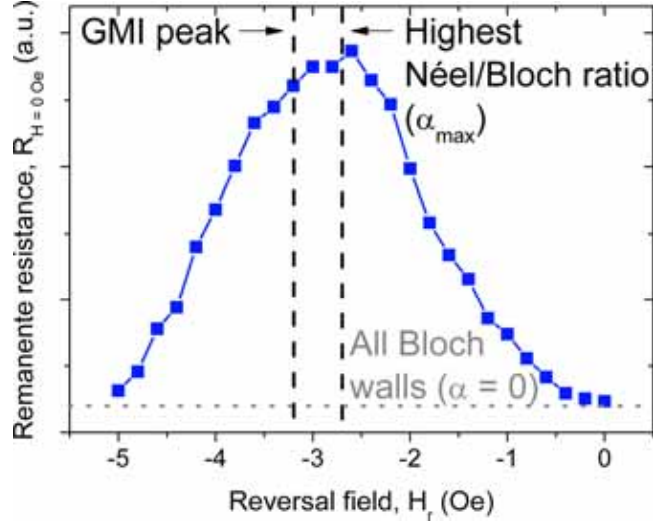

FIG. 6. Remanent GMI signal as a function of the reversal field values. The vertical dashed lines indicated the position of the GMI major curve peak and the highest Néel/Bloch wall ratio, while the horizontal dotted line shows the level for only Bloch wall type.

The $\alpha$ evolution divides the FORCs into three groups (Fig. 5). FORCs initiating below $\alpha$ maximum ($H_r < -2.7$ Oe, blue lines) move out of the major curve path around the positive peak, where they reach higher GMI signal. We suppose that it is a consequence of still remaining Bloch walls turning into Néel ones, the particular FORC path achieving to increase the $\alpha$ maximum in comparison of the major path. Following FORCs present a mix of Bloch and Néel walls (red lines) until $H_r = 0$ Oe, where no more DW transition occurs, all being of the Bloch type (green lines), which yields to a null FORC distribution.

## B. Dual-hysteron model

The complex irreversible behavior of the GMI signal requires more than one simple bistable hysteron to be adequately modeled [21]. The complete process, each associated with one DW in the ribbon and called dual hysteron, is characterized by only one saturation level, of lower $\mu_t$, associated with a Bloch DW (Fig. 7). Starting from positive saturation field, the Bloch-Néel transition, which leads to a higher $\mu_t$, occurs at a certain negative field, denoted $H_{BN}$. From this point, the further behavior depends on the applied field variation, i.e. the transition back to Bloch DW is possible for two different applied field values, one negative where $H_{NB} < H_{BN}$ and one positive, $H^+_{NB}$. If the field is increasing until $H^+_{NB}$, the DW returns back to a Bloch structure, closing a first simple two-transitions hysteron called anti-hysteron, because it yields to an anti-peak in the FORC distribution, i.e. $\rho < 0$. On the other side, if the field is sufficiently decreased after reaching $H_{BN}$, the DW undergoes the Néel-Bloch switch at $H_{NB}$. This transition allows the system to reach the negative saturation position, initiating at the same time the second simple two-transitions hysteron of the $\mu_t$ irreversible behavior. This hysteron, which gives a positive FORC distribution, presents a Bloch-Néel transition for a positive applied field, $H^+_{BN}$. Then, there are again two possibilities for the return to Bloch wall transition: at $H_{NB}$ while staying on the same hysteron, or changing to the anti-hysteron by switching at $H^+_{NB}$. In summary, the necessary applied field to provoke a Bloch to Néel transition depends on the sign of the last saturation reached. On the other side, return to Bloch wall can occur at two different field values, without influence of the applied magnetic field past history. This yields a pair of hysteron/anti-hysteron that is linked together through the Néel wall state and is called dual-hysteron.

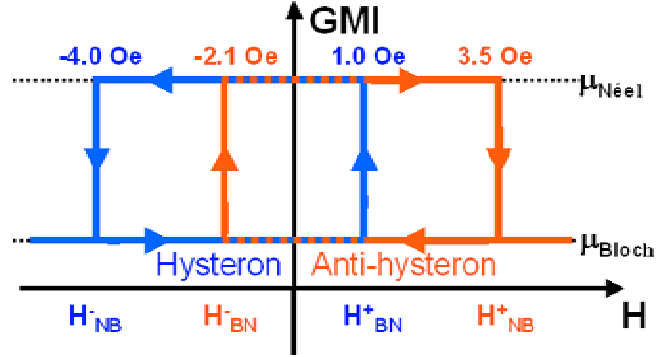

FIG. 7. Dual-hysteron model of the GMI bi-hysteretic behavior, which can be decomposed into a hysteron (blue)/anti-hysteron (red) pair. It represents the irreversible switches of the μ, generated by the DW transitions.

The FORC distribution mainly exhibits four features (positive or negative peaks), all located in low $H_r$ values, two of which may be directly associated to the hysteron/anti-hysteron pair (Fig. 8). The positive peak A corresponds to the hysteron. Its position can be considered as the average transition fields: after a Néel to Bloch transition at $H_r = H_{NB} = -4.0$ Oe, the return to Néel wall occurs at $H = H^+_{BN} = 1.0$ Oe, giving rise to a hysteron centered on $-1.5$ Oe and large of 5.0 Oe. The broad distribution is in agreement with the hypothesis of gradual DW transitions. It is worth noticing that this information can not be obtained from the major GMI spectra, because it is related to physical processes occurring locally in the soft magnetic ribbon. Obviously, the existence of this hysteron requires a former Bloch-Néel transition, as well as an ulterior switch back to Bloch wall. This last process leads to the negative peak B, located along the same $H_r$ than peak A, but for higher $H$. The $H$ distance between the two features, 1.4 Oe, corresponds to the average interval of field for which a domain wall sustains a Néel structure. The trace of the former Bloch-Néel transition is visible of the FORC distribution at $H_r = H_{BN} = -2.1$ Oe. This negative peak (C) results from the anti-hysteron path, where the return to Bloch structure occurs for $H = H^+_{NB} = 3.5$ Oe. Therefore, we can observe that the anti-hysteron is not symmetrically located (0.7 Oe) and larger (5.6 Oe) than the hysteron. The $H_{BN}$ values extracted are in agreement with the previously estimated critical field from the Bloch-Néel DW transition model ($\approx$ 2 Oe). Finally, it is possible to distinguish a fourth peak (D), positive and approximately located at $H = -H_r = 2.1$ Oe. This position is not directly related to any of the hysterons previously identified. We assume that it represents the consequence of additional Bloch-Néel transitions during the FORCs return path, which lead the FORCs to reach a higher GMI signal than the major curve. This assumption is reinforced by the vertical elongation of peak D, this phenomenon occurring for $H_{BN} < H_r < H_{NB}$.

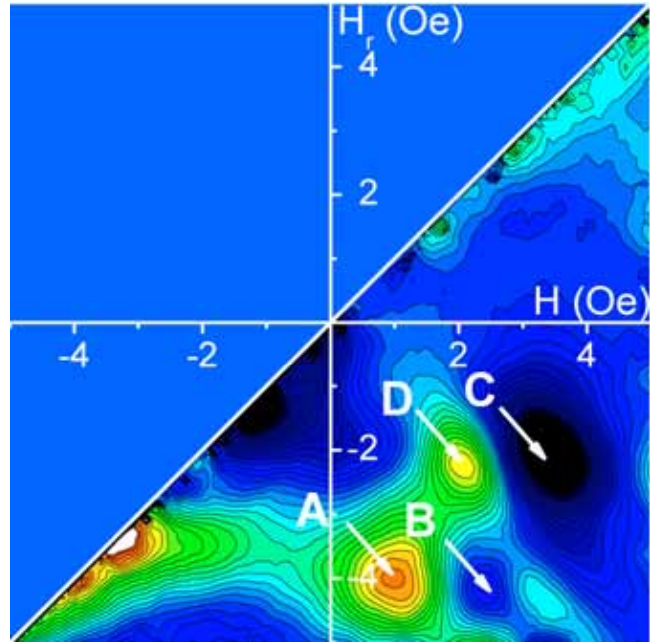

FIG. 8. Typical FORC diagram from the GMI signal of an amorphous ribbon. The positive part ranges from pale blue ($\rho = 0$) to red (maximum value of $\rho$), while blue goes darker as the $\rho$ value decreases in the negative part. It's noteworthy that the experimental data cover only the lower triangle of the contour plot, because no data are taken for $H < H_r$.

## C. Parameters influence

The real and imaginary ($X$) parts of the GMI signal present similar FORC distributions, the only perceptible difference being broader hysteron/anti-hysteron peaks for $X$, which can end up to completely cover the other features. In both cases, decreasing the ac current frequency ($f$ = 200-500 kHz) shifts the peak A toward lower $H$ (Fig. 9). From a physical point of view, it means that the Bloch to Néel transition requires a weaker longitudinal static field to be energetically favorable. Indeed, reducing $f$ increases the penetration depth, increasing thereby the ribbon volume submitted to a perpendicular field. The critical angle of the effective $M$ for the Bloch to Néel transition is therefore reached for lower values of longitudinal applied field. A similar trend is observed with the $H_{BN}$ values extracted from peak C. However, the broadness of this distribution gives less accurate results for the peak position. Finally, the magnetostriction values strongly influence the FORC distribution (Fig. 9), which is expected because it modifies the anisotropy constant and therefore the critical field for the DW transition. A detailed study of the magnetostriction effect on the GMI hysteretic behavior, done by FORC method, is currently under progress.

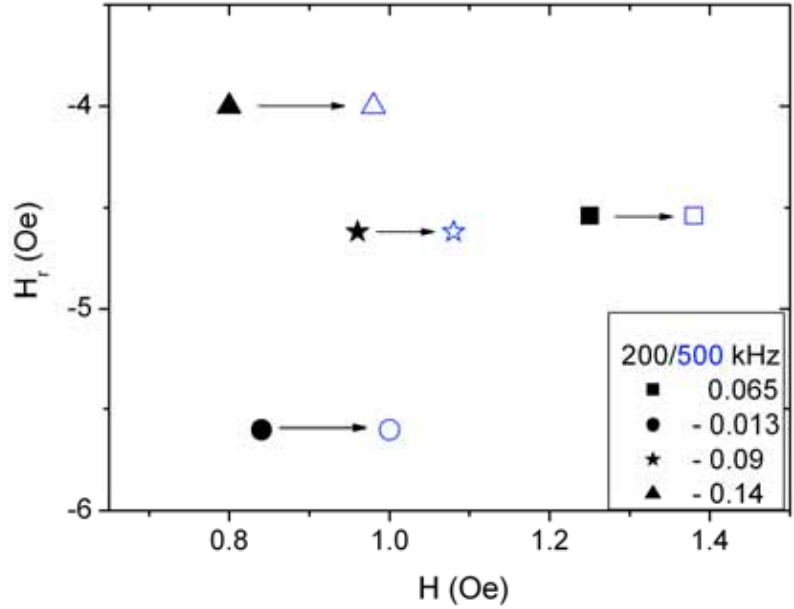

FIG. 9. Position of the maximum of the peak A on the real part FORC distribution, in function of the magnetostriction constant and the frequency measurement.

**V. CONCLUSION**

In conclusion, we have applied the FORC formalism to study the hysteretic behavior of GMI effect and attributed it to $\mu_t$ irreversibility created by DW transitions. It's important to point out the fully reversible character of the corresponding *MH* curve. We propose a dual-hysteron model, which successfully explains the resulting FORC distribution. This method now gives us the tools to adequately understand this GMI peculiarity, which is important for both fundamental magnetic studies and technological applications. The approach presented here opens the possibility for further studies concerning the FORC analysis applied to different hysteretic systems through magneto-transport phenomena.

## ACKNOWLEDGMENTS

This work was financially supported in part by the Brazilian agencies FAPESP, FAPEMIG, CAPES and CNPq and Canadian agency FQRNT. J. M. Barandiaran is acknowledged for supplying the samples. A. Yelon and D. Ménard are also acknowledged for fruitful ideas and discussions regarding GMI in novel magnetic materials.